\documentclass[aps,prb,preprint,groupedaddress,showpacs,amsmath,amssymb]{revtex4}

\usepackage{graphicx}
\usepackage{dcolumn}
\usepackage{bm}

\begin{document}

\title{Visualization of ferromagnetic domains in TbNi$_2$B$_2$C and ErNi$_2$B$_2$C single crystals: Weak ferromagnetism and its coexistence with superconductivity.}

\author{I.S. Veschunov, L.Ya. Vinnikov}
\affiliation{Institute of Solid State Physics RAS, Chernogolovka, Moscow region 142432, Russian Federation}
\author{S.L. Bud'ko, P.C. Canfield}
\affiliation{Ames Laboratory US DOE and Department of Physics
and Astronomy, Iowa State University, Ames, IA 50011, USA}

\date{\today}

\begin{abstract}

The magnetic flux structure in the basal plane, (001), of single crystals of superconducting (R = Er) and
non-superconducting (R = Tb) RNi$_2$B$_2$C was studied by high resolution Bitter decoration at temperatures below
$T_c$ (superconducting transition) and/or $T_N$ (antiferromagnetic transition). For both materials two sets of
domain boundaries, in \{110\} and \{100\} planes, were observed. The temperature ranges in which the \{100\}
domain boundaries were observed in TbNi$_2$B$_2$C and ErNi$_2$B$_2$C coincide with the weak ferromagnetic (WFM)
ordering in these materials. On the other hand, the \{110\} twin boundaries - the antiferromagnetic domain
boundaries - were observed in both compounds below $T_N$. The possibility of interpretation of \{100\} boundaries
as Bloch domain walls in the weakly ferromagnetic phase, for $T < T_{WFM} < T_N$ (TbNi$_2$B$_2$C) or $T < T_{WFM}
< T_N < T_c$ (ErNi$_2$B$_2$C) is discussed.

\end{abstract}


\maketitle

The coexistence of superconductivity and magnetism is studied in two distinct classes of materials: artificially
structured (multilayered) systems \cite{lyu05a,buz05a,rya03a,rya01a} and spatially uniform (bulk) crystals. The
coexistence of ferromagnetism and superconductivity in bulk materials has been observed in a very limited number
of cases.  In the RRh$_4$B$_4$ series the temperature interval of coexistence of superconductivity and
ferromagnetism is rather narrow \cite{buz84a,fis90a}. For superconducting ferromagnets with $T_c < T_{FM}$
(rutheno-cuprates, uranium compounds under pressure) a variety of domain structures has been predicted
\cite{fau05a,son02a}. Over the past decade, the family of rare earth borocarbides, RNi$_2$B$_2$C (R = rare earth)
\cite{cav94a} has become a convenient model system in which long range magnetic order (associated with magnetic
rare earth ions \cite{lyn97a}), superconductivity and the coexistence of these two phenomena is observed with the
specific ground state depending on the specific rare earth ion \cite{can98a,mul01a,mul02a,bud06a}. For example,
for R = Y, Lu, superconductivity with $T_c \approx 15.5$ K and 16.5 K, respectively, was observed, while
GdNi$_2$B$_2$C has an antiferromagnetic transition at $T_N \approx 20$ K and no superconductivity (down to 30 mK).
In ErNi$_2$B$_2$C superconductivity ($T_c \approx 10.5$ K) coexists with magnetic long range order:
antiferromagnetism, AFM, ($T_N \approx 6$ K) and weak ferromagnetism, WFM, ($T_{WFM} \approx 2.3$ K)
\cite{can96a}. It is important to mention that in the AFM and WFM phases the magnetic moment is confined to the
basal plane, (001), with the easy axis along $[100]$ or $[010]$ direction \cite{cho01a,kaw02a}. For TbNi$_2$B$_2$C
two magnetic phases, AFM ($T_N \approx 15$ K) and WFM ($T_{WFM} \approx 8$ K) were reported \cite{cho96a,der96a}.

The magnetic flux structures in RNi$_2$B$_2$C (R = Er, Ho, Tb) single crystals (in a limited range of temperatures
and magnetic fields) were recently studied by high resolution Bitter decoration technique
\cite{sah00a,vin03a,vin05a}. For ErNi$_2$B$_2$C and HoNi$_2$B$_2$C antiferromagnetic domains induced by
magnetoelastic interactions were observed \cite{sah00a,vin05a}. In these materials, vortex pinning was detected on
the \{110\} twin AFM domain boundaries in the orthorhombic phase. Unexpectedly, magnetic contrast on similar,
\{110\} - type boundaries was also observed in non-superconducting TbNi$_2$B$_2$C \cite{vin03a}. The nature of
this magnetic contrast remains unclear. Additionally, in some regions of the (001) surface, a more complex pattern
was observed: in addition to the dominant stripes along \{110\} boundaries, a  fine structure on $[100]$ and
$[010]$ boundaries was observed (see Fig. 3c in Ref. \onlinecite{vin03a}).
\\

In this work, to elucidate the nature of the domain structure associated with the AFM and WFM phases,
TbNi$_2$B$_2$C and ErNi$_2$B$_2$C single crystals were studied by high resolution Bitter decoration at
temperatures that extend the lower limit of this technique. This is done so as to study the WFM phase of
ErNi$_2$B$_2$C at $T \sim T_{WFM} \approx 2.3$ K. These compounds are similar in their crystal structure and
magnetic order, in particular they have a transition to WFM state below the respective Neel temperature, and
therefore are suitable for a comparative study. We observed characteristic features of magnetic contrast in basal
(001) plane along $<100>$ and $<110>$ directions in non-superconducting TbNi$_2$B$_2$C crystals. In addition,
preferential accumulation of vortices in (100) and (010) planes, distinct from the pinning by \{110\} - type twin
boundaries, were observed for the first time in superconducting ErNi$_2$B$_2$C single crystals just above $T_{WFM}
= 2.3$ K.
\\

The Bitter decoration technique is based on the segregation of fine dispersed magnetic particles on the surface of
a superconductor or a magnetic material due to the regions of inhomogeneous magnetic flux \cite{tra66a,vin93a}.
Scanning electron microscope images of the magnetic particle distribution (bright dots in the figures below)
provide information about the structure of vortices or magnetic domains. The high resolution ($< 100$ nm) of this
technique is due to the small size of the magnetic particles ($< 10$ nm) that are produced, at low temperature, by
evaporation of the magnetic material (in our case, iron) from the surface of a tungsten wire in a buffer gas
(helium) atmosphere at low pressure ($\sim 10^{-2}$ Torr) \cite{vin93a}.

TbNi$_2$B$_2$C and ErNi$_2$B$_2$C single crystals were grown using a Ni$_2$B, high temperature flux method
\cite{can98a,mxu94a}. As grown (001) surfaces with approximate dimensions of $3 \times 5 $ mm$^2$ were used in
these experiments. The decoration was performed in the field-cooled regime with the magnetic field $\sim 20$ Oe
for ErNi$_2$B$_2$C and $\sim 1.1$ kOe for TbNi$_2$B$_2$C, applied along the $c$ axis  of the crystals (this 50 -
fold difference in applied field is associated with the fact that in ErNi$_2$B$_2$C the vortices provide the flux
contrast whereas in TbNi$_2$B$_2$C the contrast has to come directly from the local moment structure).

The decoration temperature, $T_d$, was in the range of 2.6 K to 15 K. It should be mentioned that in these
experiments the decoration temperature is defined as the temperature of the gas measured by a resistive
thermometer, placed in the vicinity of the sample, just after the current through the evaporator is turned on.
During the thermal evaporation of iron the heating of the system occurs so that the temperature in the evaporation
chamber may increase by several degrees, therefore in the experiment the decoration chamber is initially cooled
down to the temperature $T_1$ ($T_1 < T_d$). The decoration temperature, $T_d$, is minimized by decrease of the
size (mass) of the evaporator using a 0.05 mm diameter tungsten wire. The lowest $T_d$ in decoration experiments
reported in the literature so far is 2.9 K \cite{mar97a}. In our experiments the lowest temperature $T_1$ was 1.6
K, in this case the $T_d$ was estimated as 2.6 K. These reduced temperatures allow us to try to image the WFM
state in ErNi$_2$B$_2$C which has $T_1 < (T_{WFM} \approx 2.3$ K) $\leq T_d$.

Scanning electron microscope (SEM) was used for visualization of the decoration patterns after warming of the
sample up to the room temperature.
\\

\begin{figure}
\begin{center}
\includegraphics[angle=0,width=120mm]{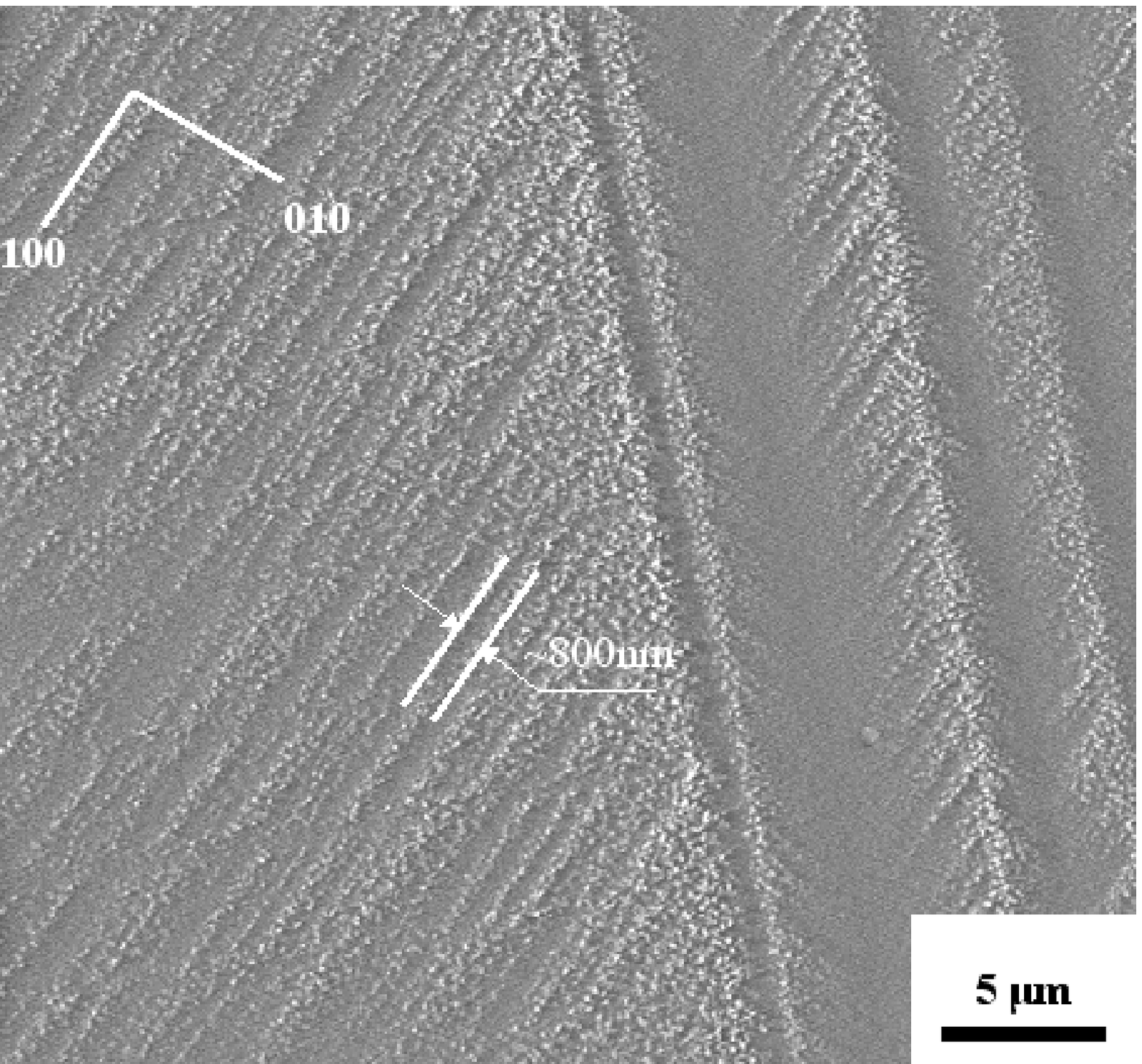}
\end{center}
\caption{Bitter decoration pattern on a TbNi$_2$B$_2$C single crystal cooled down to $T_1 = 4.2$ K in a magnetic
field of $H = 1.1$ kOe and decorated at $T_d \approx 7$ K.}\label{f1}
\end{figure}

A Bitter decoration pattern on the (001) surface of TbNi$_2$B$_2$C single crystal after field cooling in a field
of $H = 1.1$ kOe applied along the $c$ axis is shown in Fig. \ref{f1}. One type of domain boundaries are aligned
along the $<110>$  directions. These are often separated by 3-10 $\mu$m, but (as can be seen in Fig. \ref{f1}) can
also have much larger spacings as well. A second set of boundaries, aligned along the $<100>$ directions appear
with an order of magnitude smaller separation between them. Similar structures were observed in TbNi$_2$B$_2$C for
temperatures $T_d$ below 8 K. Graphically similar patterns (regions with two distinct sets of boundaries) were
observed in other parts of the sample and in different samples. A detailed discussion of the magnetic flux
structures observed in TbNi$_2$B$_2$C single crystals in a wide range of temperatures and magnetic fields will be
presented elsewhere \cite{ves07a}.

The symmetry of the TbNi$_2$B$_2$C crystal structure allows for formation of AFM structures with a weak
ferromagnetic component in the basal, $(ab)$ plane \cite{cho96a,der96a}. The direction of the weak ferromagnetic
moment coincides with the direction of the longitudinally polarized spin density wave along $[100]$ or $[010]$.
The temperature range in which the fine structure in decoration patterns is observed along these directions
coincides with the temperatures at which WFM phase is observed in TbNi$_2$B$_2$C. It is natural then to suppose
that the observed fine structure corresponds to the boundaries of the WFM domains, i.e. these boundaries cause the
appearance of the magnetic contrast. Since the Bitter decoration technique is only sensitive to the component of
the inhomogeneous magnetic field perpendicular to the surface, the observed domain boundaries are most probably of
the Bloch type. It is possible that the boundaries between the AFM domains with mutually perpendicular directions
of the spin density wave contribute to the magnetic contrast as well \cite{vin03a}.

\begin{figure}
\begin{center}
\includegraphics[angle=0,width=120mm]{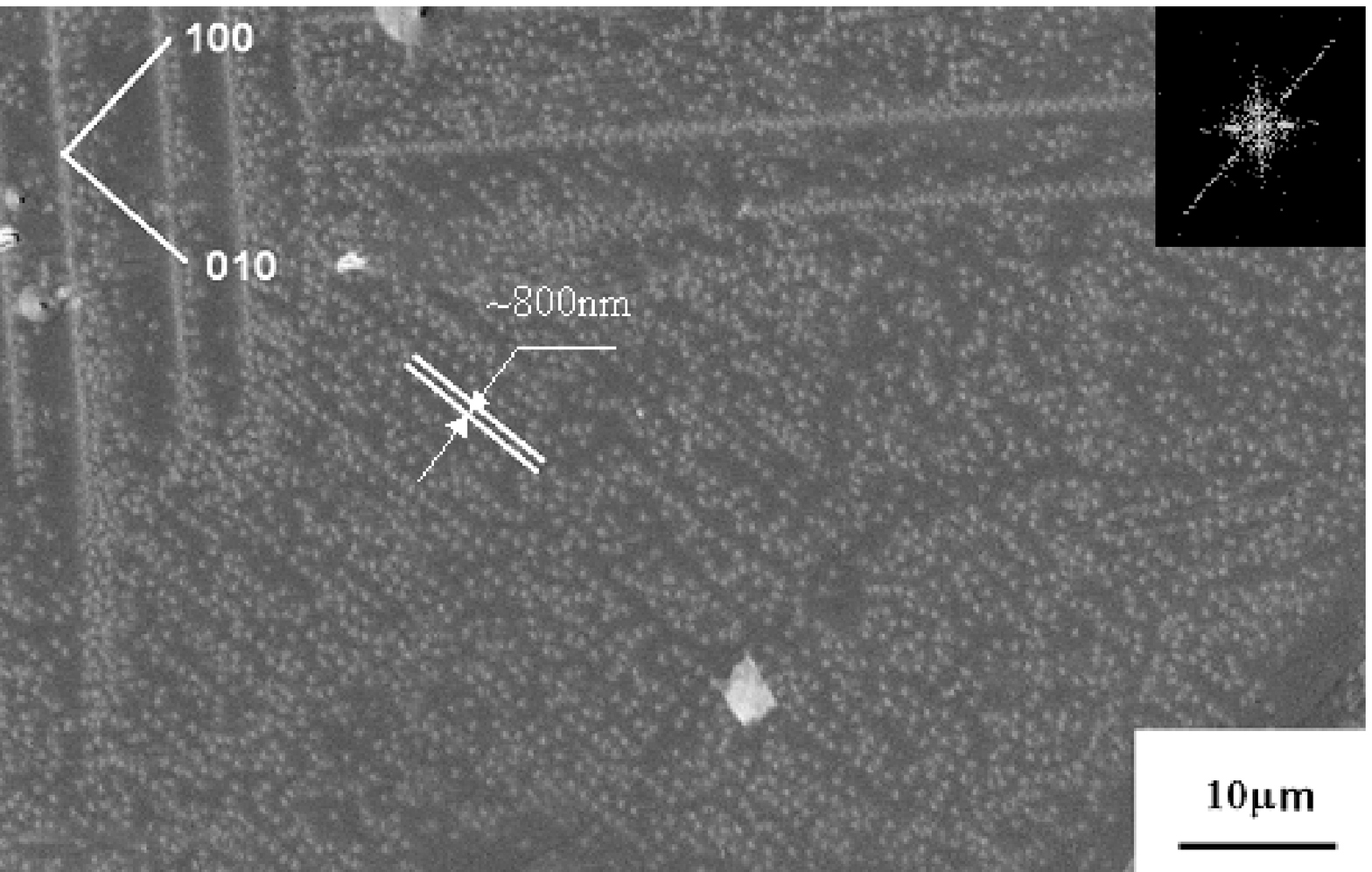}
\end{center}
\caption{Bitter decoration pattern on a twin-boundaries-free region of a ErNi$_2$B$_2$C single crystal cooled down
to $T_1 = 1.8$ K in a magnetic field of $H = 20$ Oe and decorated at $T_d \approx 4$ K. Inset: Fourier transform
of the image.}\label{f2}
\end{figure}

For this work it is important to note that TbNi$_2$B$_2$C is magnetically similar to ErNi$_2$B$_2$C: similar local
moment anisotropy and similar antiferromagnetic order that leads to similar weak ferromagnetism at lower
temperatures. The key differences between ErNi$_2$B$_2$C and TbNi$_2$B$_2$C are that $T_N$ and $T_{WFM}$ for
ErNi$_2$B$_2$C are substantially lower than those for TbNi$_2$B$_2$C and ErNi$_2$B$_2$C superconducts below $T_c
\approx 10.5$ K. This means that in ErNi$_2$B$_2$C the interaction of vortices with magnetic order can be probed.
In particular, if the base temperature for decoration, $T_1$, is below $T_{WFM} = 2.3$ K and the decoration
temperature, $T_d$, only slightly exceeds $T_{WFM}$, then in addition to vortices pinned along the AFM domain
boundaries in the \{110\} planes, \cite{sah00a,vin05a} other features may be observed. Indeed, rows of vortices
along the $<100>$ directions were detected in several regions (where the twin boundaries were absent) of the
ErNi$_2$B$_2$C crystal (Fig. \ref{f2}). A Fourier transform of this pattern (shown in the inset to Fig. \ref{f2})
confirms the alignment of the rows of vortices along $[010]$ (seen as a line in $[100]$ direction in the insert)
and allows for the estimate of the minimum distance between the rows of vortices to be of the order of several
hundreds nanometers.

Although the data presented in figure \ref{f2} are compelling it does have to be pointed out that these data have
been collected at the edge of what the Bitter decoration technique can achieve.  It should again be noted that
whereas $T_1$ is below $T_{WFM}$, $T_d$ is above this key temperature.  This makes the decoration a dynamic
process:  e.g. one that takes place as the sample is warming toward $T_{WFM}$ from below.   In addition the data
have been collected at the edge of what the ErNi$_2$B$_2$C samples allow in terms of resolvability and
reproducibility. The pinning of the vortices along the $<100>$ directions is only clearly seen in regions of the
sample where the \{110\} twin boundaries are absent.  The details of the twin boundary spacing appear to be sample
dependent and are probably are complex functions of multiple extrinsic, strain effects.

\begin{figure}
\begin{center}
\includegraphics[angle=0,width=120mm]{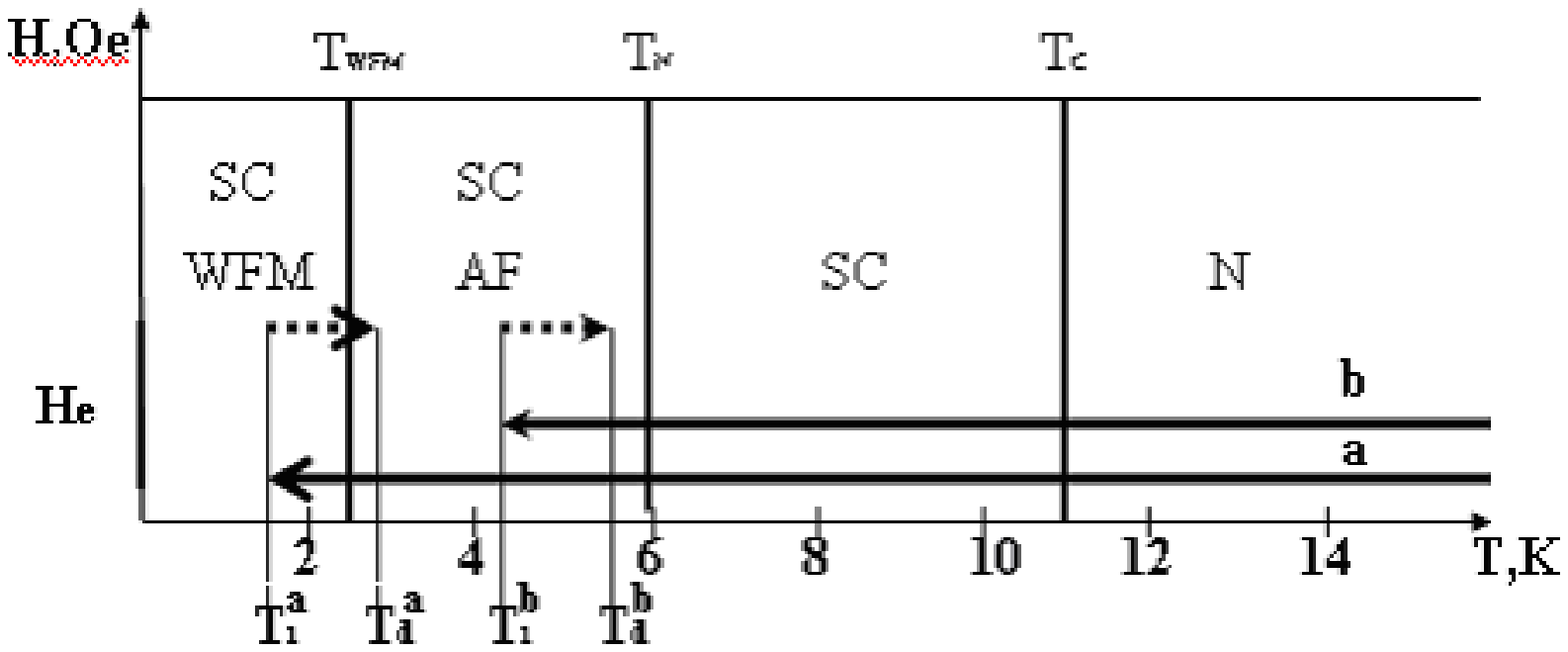}
\end{center}
\caption{Illustration of the protocols used in the low temperature experiment. Path (a): cooling down to $T_1$
below $T_{WFM}$, $T_d$ slightly above $T_{WFM}$; path (b): both $T_1$ and $T_d$ above $T_{WFM}$ but below
$T_N$.}\label{f3}
\end{figure}

\begin{figure}
\begin{center}
\includegraphics[angle=0,width=90mm]{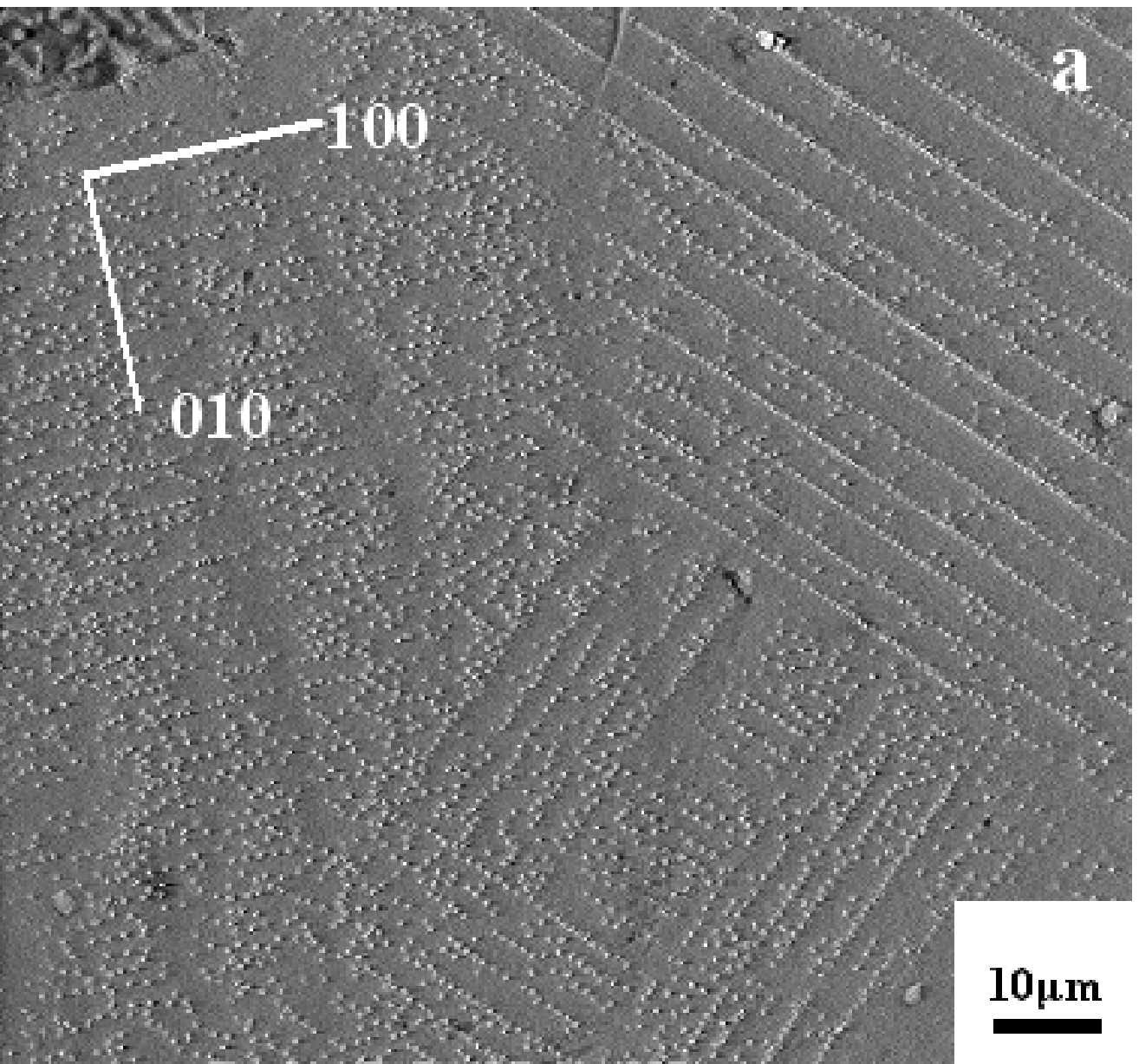}
\includegraphics[angle=0,width=90mm]{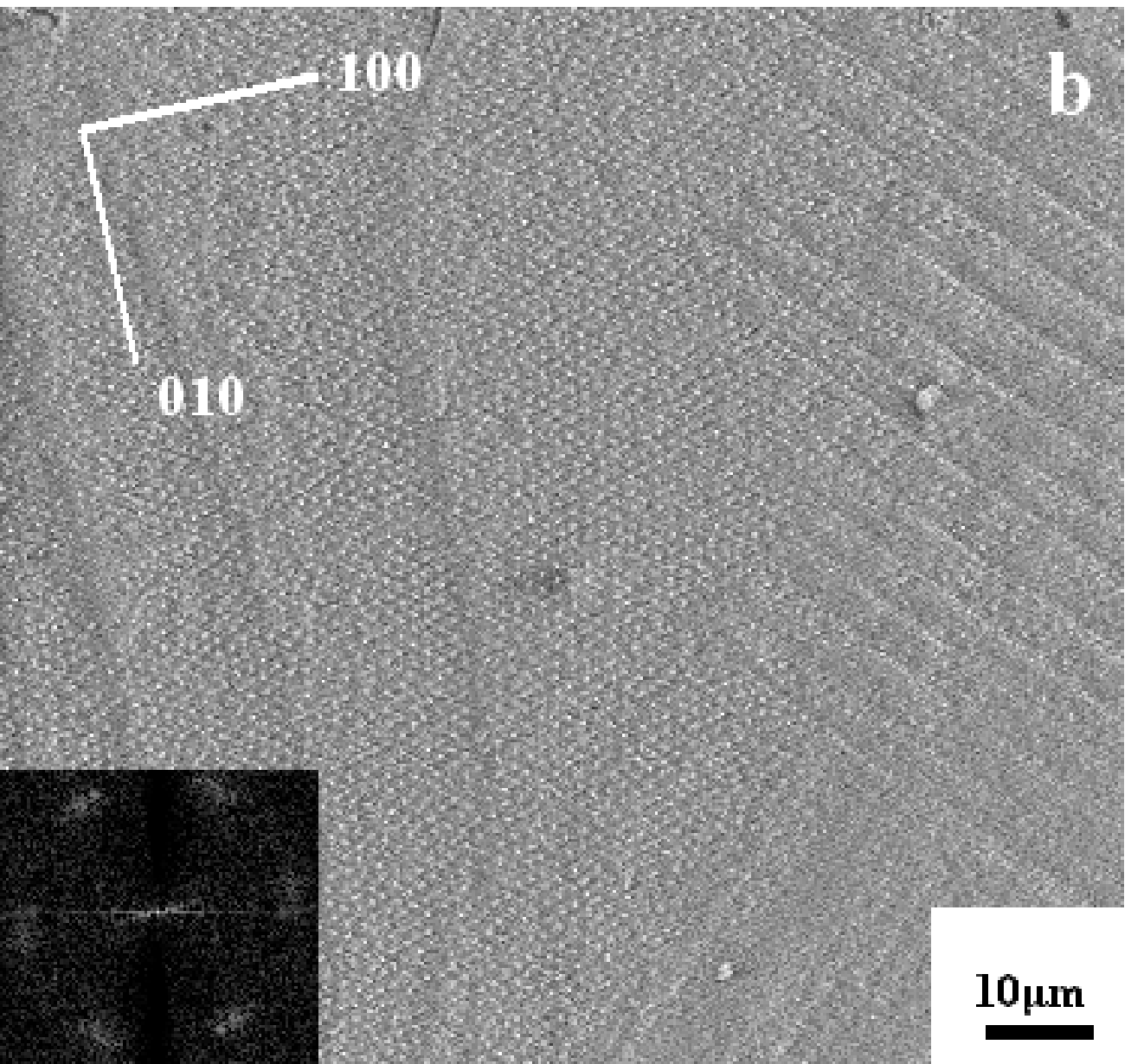}
\end{center}
\caption{Bitter decoration pattern on an ErNi$_2$B$_2$C single crystal (sample I) obtained after field cooling in
an applied field $H$, using experimental protocols (a): $T_1 = 1.8$ K, $T_d = 2.6$ K, $H = 15$ Oe (panel (a)) and
(b): $T_1 = 4.2$ K, $T_d = 5.5$ K, $H = 20$ Oe (panel b). Inset to panel (b): Fourier transform of the image from
the left side of the pattern showing hexagonal order of vortices.}\label{f4}
\end{figure}

In order to (i) more fully detail the effects of WFM on vortex state and (ii) illustrate the difficulty of these
experiments, a further set of experiments followed the evolution of the patterns observed on the same region of
the (001) surface of a ErNi$_2$B$_2$C crystal that was cooled down to different temperatures: (a) $T_1 < T_{WFM}$
and (b) $T_{WFM} < T_1 < T_N$ before decoration. The protocol of the experiments is shown schematically in Fig.
\ref{f3}. Two distinct scenarios were observed. In the first case (scenario I) after the crystal was cooled down
to $T_1 < T_{WFM} \approx 2.3$ K \cite{cho01a,kaw02a} and decorated at $T_d \geq T_{WFM}$ rows of vortices along
$[100]$ were observed, Fig. \ref{f4}a. These rows could be easily seen on the left side of the pattern. The second
image (Fig. \ref{f4}b) shows a decoration pattern of the same single crystal (after the magnetic particles from
the first decoration were carefully removed) obtained in different conditions: the crystal was cooled down to $T_1
= 4.2$ K $> T_{WFM}$ and decoration temperature was $T_d = 5.5$ K $< T_N \approx 6$ K. The left side of the image
(where no $<110>$ twin boundaries are present) shows triangular vortex lattice without a preferential orientation
of the close-packed directions of the lattice along $[100]$. The patterns shown in Figs. \ref{f4}a,b are
reproducible on subsequent decoration experiments after cycling of the sample to room temperature.

\begin{figure}
\begin{center}
\includegraphics[angle=0,width=90mm]{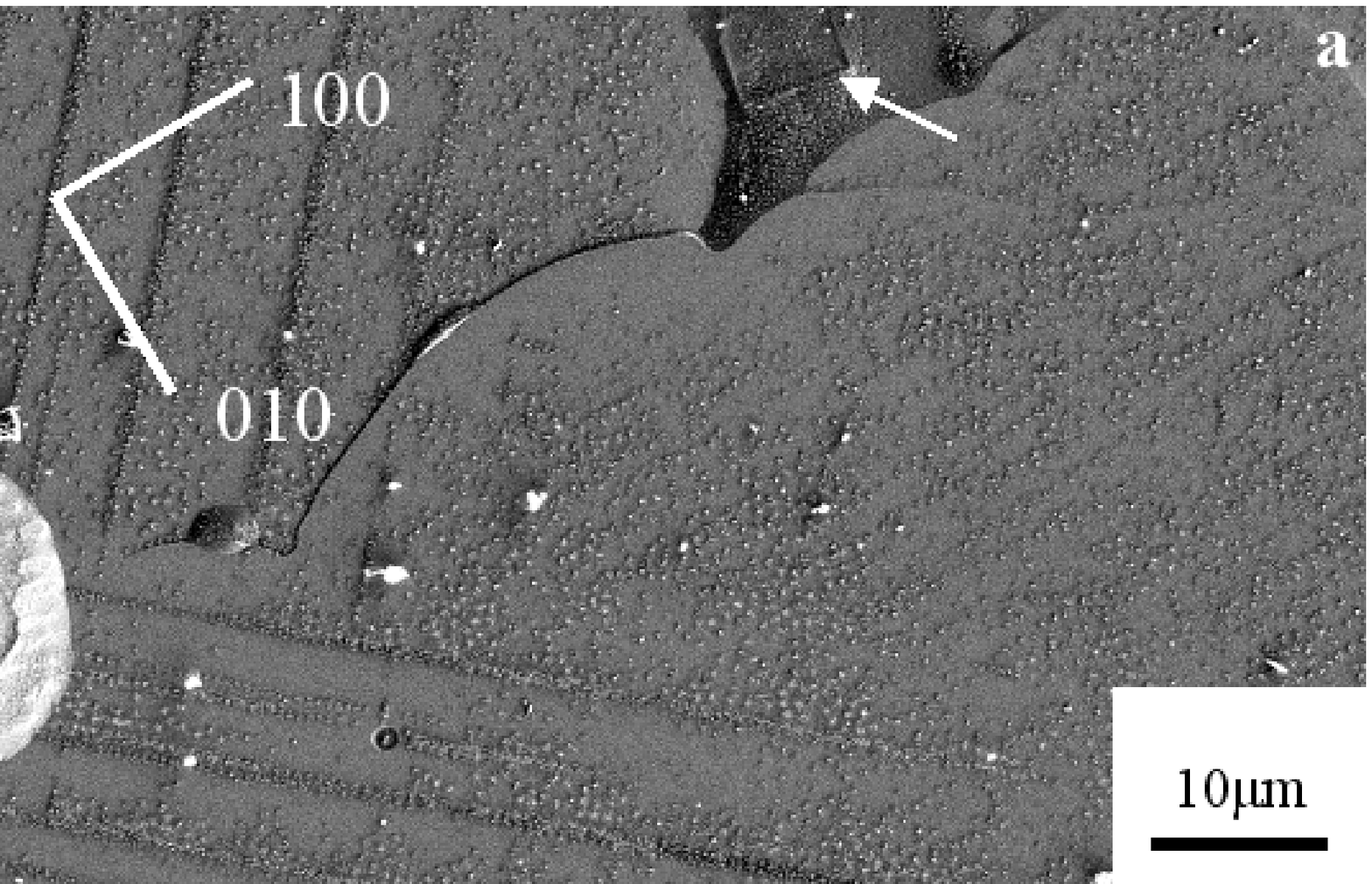}
\includegraphics[angle=0,width=90mm]{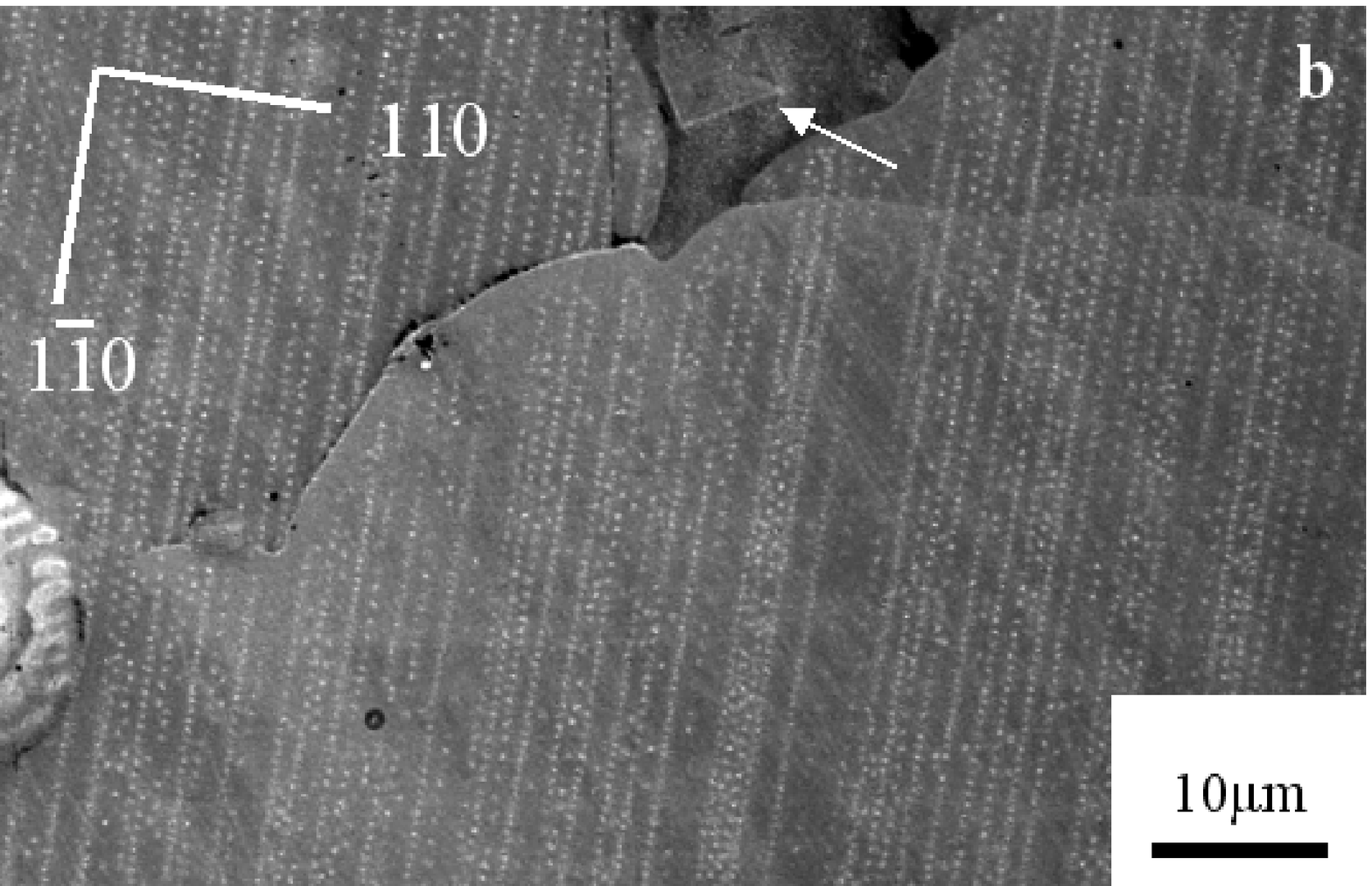}
\end{center}
\caption{Bitter decoration pattern on an ErNi$_2$B$_2$C single crystal (sample II) obtained after field cooling in
applied field $H$, using experimental protocols (a): $T_1 = 1.6$ K, $T_d = 4.2$ K, $H = 20$ Oe (panel (a)) and
(b): $T_1 = 4.2$ K, $T_d = 5.5$ K, $H = 20$ Oe (panel b).}\label{f5}
\end{figure}

In the second case (scenario II), in decoration experiments on a different ErNi$_2$B$_2$C crystal, the first
cooling down to $T_1 < T_{WFM}$ and decoration at $T_d \geq T_{WFM}$ resulted in the pattern shown in Fig.
\ref{f5}a, with rows of vortices along $<100>$ directions observed in several parts of the crystal where the twin
boundaries along \{110\} were absent. On a subsequent cooling down to $T_1 = 4.2$ K $> T_{WFM}$ and decoration
below $T_N$ only rows of vortices along twin boundaries \{110\} were observed, including the regions of the
crystal where they were absent on initial decoration experiment (Fig. \ref{f5}b). It is important to mention that
only patterns similar to one in Fig. \ref{f5}b were observed in further experiments on this crystal.

Out of six ErNi$_2$B$_2$C crystals studied, scenario I was realized in two, scenario II was observed in one, and
the other three crystals showed only the \{110\} twin boundaries.

In all cases they were observed, the distance between the rows of vortices along the $<100>$ direction in
ErNi$_2$B$_2$C is much smaller that between the the rows of vortices along the $<110>$ (Figs.
\ref{f2},\ref{f4},\ref{f5}), similar to the distances between alleged magnetic domain boundaries in TbNi$_2$B$_2$C
crystals (Fig. \ref{f1}).
\\

Our observations in ErNi$_2$B$_2$C crystals can be interpreted as a visualization of the pinning of vortices on
the WFM domains boundaries in (100) or (010) planes. Since we did not succeed in lowering $T_d$ for ErNi$_2$B$_2$C
crystals below $T_{WFM} = 2.3$ K, we possibly observe the remnant pinning of vortices at $T_d$ slightly above
$T_{WFM}$ that does not vanish due to kinetics reasons. It is plausible that the vortices forming the rows
"freeze" at the domain boundaries of the low temperature WFM phase, thus visualizing the domain boundaries in the
(100) or (010) plains. The distance between the rows of vortices is probably not exactly the distance $l$ between
the WFM domain boundaries due to the repulsive interaction between vortices that exists even in small ($< 100$ Oe)
magnetic fields. From Figs. \ref{f2},\ref{f4},\ref{f5} the reasonable estimate of $l$ seems to be $l < 1$ $\mu$m.
The width of the domain wall, $w$, and the width of the domain, $l$ can be estimated following the theoretical
model \cite{fau05a} for domain structure in superconducting ferromagnets: if such magnetic domains exist, the
width of the domain wall should be less than superconducting penetration depth, $\lambda$. If we take the size of
the AFM domain, $d$, as an effective size of the region of the AFM to WFM transformation, and use the condition $w
\leq \lambda$, then $l \approx \sqrt{dw}$. For ErNi$_2$B$_2$C, using $\lambda = 70$ nm \cite{blu06a} and $d
\approx 7$ $\mu$m, $l$ is estimated as $\approx 700$ nm, similar to the observed distances between the rows of
vortices along $<100>$ (Figs. \ref{f2},\ref{f4},\ref{f5}). It should be stressed that although in both the
ErNi$_2$B$_2$C and TbNi$_2$B$_2$C cases we are able to visualize WFM domain boundaries, in TbNi$_2$B$_2$C the
magnetic particles image directly the magnetic flux gradients on the domain boundaries, whereas in ErNi$_2$B$_2$C
we use magnetic flux gradient of superconducting vortices that interact with the magnetic domain boundaries as
well as with each other. Thus the "texture" of the images pertaining to ErNi$_2$B$_2$C and TbNi$_2$B$_2$C is
different.

AFM to WFM transition in TbNi$_2$B$_2$C and ErNi$_2$B$_2$C was discussed as a lock-in phase transformation from
incommensurate to commensurate spin density wave \cite{wal03a}. Neutron scattering data \cite{wal03a} suggest that
the domain size in WFM phase of ErNi$_2$B$_2$C is of the order of 1000$a$, where $a \approx 0.35$ nm is the
lattice parameter of ErNi$_2$B$_2$C. This yields an estimate value $l \approx 350$ nm, that is in a reasonable
agreement with our results. Additionally, recently \cite{bre06a} the characteristic dimensions of the AFM domain
structure in orthorhombic phase of ErNi$_2$B$_2$C, similar to YBa$_2$Cu$_3$O$_7$ were estimated considering
minimum of the elastic energy (WFM phase was not considered), the results are consistent with the picture
presented here.

There were several attempts \cite{blu06a,kaw01a} to observe spontaneous flux line lattice in the WFM phase ($T <
T_{WFM} < T_c$) of ErNi$_2$B$_2$C. In neutron experiments \cite{kaw01a} no spontaneous flux line lattice was
observed in zero field cooled protocol, however a vortex phase was observed after magnetic field was applied below
$T_{WFM}$, that is consistent with our observations that changes in decoration patterns were observed only if the
sample was cooled in an applied field below $T_{WFM}$. Hall microscope measurements \cite{blu06a} were not able to
resolve the structure of magnetic flux on (100) or (010) planes. An estimate of the spontaneous magnetization in
the basal plane give the value of the in-plane magnetic induction $\sim 700$ G, that corresponds to the period of
$\sim 0.16$ $\mu$m, below the resolution of the Hall microscope \cite{blu06a}, for the anticipated spontaneous
flux line lattice in (100) or (010) planes.
\\

{\it In summary}, we observed (a) residual pinning of vortices  on (100) or (010) boundaries in ErNi$_2$B$_2$C
slightly above $T_{WFM} = 2.3$ K that visualize the WFM domains in this material; (b) magnetic contrast on (100)
or (010) boundaries (that possibly are Bloch domain walls) in WFM phase of TbNi$_2$B$_2$C, below $T_{WFM}= 8$ K.

The correspondence of the temperature intervals in which particular features in magnetic flux structure in
TbNi$_2$B$_2$C and ErNi$_2$B$_2$C were observed with the (respective) temperature ranges of WFM ordering, as well
as the orientation of these features with respect to crystallographic axis of the crystals give, in our opinion,
strong support to our interpretation of the low temperature features in decoration patterns. However, the
structure of domain walls and the nature of the magnetic contrast at \{110\} AFM domains in TbNi$_2$B$_2$C as well
as possibility of formation and visualization of spontaneous flux line lattice in the WFM phase of ErNi$_2$B$_2$C
remain as open problems and require further studies.

\begin{acknowledgments}
The authors thank D.V. Matveyev for his assistance in SEM observations and G.V. Strukov for help in preparation of
evaporators. This work was performed within the framework of the program of fundamental research "Effect of
atomic, crystalline and electronic structures on properties of condensed matter" of the Division of Physical
Sciences, Russian Academy of Sciences and Project RFFI-06-02-72025. Work at Ames Laboratory was supported by the
US Department of Energy - Basic Energy Sciences under Contract No. DE-AC02-07CH11358.
\end{acknowledgments}

\clearpage

\clearpage

\clearpage

\clearpage

\clearpage

\end{document}